\documentclass[aps,pra,twocolumn,showpacs,superscriptaddress,letterpaper,longbibliography]{revtex4-1}

\usepackage{graphicx}
\usepackage{mdframed}
\usepackage{framed}
\usepackage{indentfirst}
\usepackage{natbib}
\usepackage{amsmath,amssymb}
\usepackage{subfigure}
\usepackage{enumitem}
\usepackage{epstopdf}
\usepackage{xfrac}
\usepackage{multirow}

\begin{document}

\title{Open-ended versus guided laboratory activities:\\Impact on students' beliefs about experimental physics}


\author{Bethany R. Wilcox}
\affiliation{Department of Physics, University of Colorado, 390 UCB, Boulder, CO 80309}

\author{H. J. Lewandowski}
\affiliation{Department of Physics, University of Colorado, 390 UCB, Boulder, CO 80309}
\affiliation{JILA, National Institute of Standards and Technology and University of Colorado, Boulder, CO 80309}

\begin{abstract}
Improving students' understanding of the nature of experimental physics is often an explicit or implicit goal of undergraduate laboratory physics courses.  However, lab activities in traditional lab courses are typically characterized by highly structured, guided labs that often do not require or encourage students to engage authentically in the process of experimental physics.  Alternatively, open-ended laboratory activities can provide a more authentic learning environment by, for example, allowing students to exercise greater autonomy in what and how physical phenomena are investigated.  Engaging in authentic practices may be a critical part of improving students' beliefs around the nature of experimental physics.  Here, we investigate the impact of open-ended activities in undergraduate lab courses on students' epistemologies and expectations about the nature of experimental physics, as well as their confidence and affect, as measured by the Colorado Learning Attitudes about Science Survey for Experimental Physics (E-CLASS).  Using a national data set of student responses to the E-CLASS, we find that the inclusion of some open-ended lab activities in a lab course correlates with more expert-like postinstruction responses relative to courses that include only traditional guided lab activities.  This finding holds when examining postinstruction E-CLASS scores while controlling for the variance associated with preinstruction scores, course level, student major, and student gender.  
\end{abstract}

\maketitle

\section{\label{sec:intro}Introduction}

Laboratory physics courses represent an important and unique component of the undergraduate physics curriculum \cite{trumper2003labs}.  These courses can provide students with opportunities to engage in authentic scientific practices, develop technical lab skills, and engage collaboratively with other students.  Moreover, lab courses can be structured in such a way as to allow for considerable student autonomy when selecting interesting phenomena to investigate, designing experimental apparatus, and choosing analysis methods.  Consistent with this wide range of potential opportunities that a lab course can offer, the learning goals of lab courses often extend beyond just content delivery \cite{zwickl2013adlab}.  For example, increasing students' appreciation for, and understanding of, the nature of experimental physics has been consistently cited as an important learning outcome for laboratory courses \cite{zwickl2013adlab,trumper2003labs,AAPT2015guidelines}.  


Many undergraduate lab courses are currently taught using only traditional guided lab activities, which are typically characterized by students completing an experiment on a predetermined topic using a lab manual that guides them through the required setup, data collection, and analysis.  While there is considerable variability in these types of activities, more often than not traditional guided labs are highly structured and cookbook in nature.  These cookbook labs have been heavily critiqued as being rote and inauthentic to the process of experimental physics \cite{wieman2015labs,nap2013nrc}.  However, inauthentic lab activities can stand in opposition to the non-content goals of lab courses, such as helping students to appreciate and understand the nature and importance of experimental physics.  

In response to these and other critiques of traditional lab courses, members of the PER community have developed several new pedagogical approaches for the introductory level specifically designed to allow students to engage more authentically in the process of experimental physics.  Examples of these environments include the Investigative Science Learning Environment (ISLE) \cite{etkina2001isle}, Modeling Instruction \cite{wells1995modeling}, and integrated lab/lecture environments such as studio physics \cite{wilson1994studio} or SCALE-UP (Student-Centered Activities for Large Enrollment University Physics) \cite{beichner2000scaleup}.  A consistent feature of these pedagogical approaches is the inclusion of open-ended activities in which students have greater autonomy in what and how physical phenomena are investigated, rather than simply following instructions in a lab guide.  All of these instructional approaches were either designed with the explicit goal of improving students epistemologies about the nature of science \cite{etkina2001isle, beichner2007scaleup}, or have resulted in documented improvements in students' epistemologies, attitudes, and beliefs as measured by assessments like the CLASS (Colorado Learning Attutides about Science Survey \cite{adams2006class}) or MPEX (Maryland Physics Expectation Survey \cite{redish1998mpex}) \cite{brewe2009modeling,kohl2012studio}.  

The literature described above suggests that transformed instructional approaches that include open-ended lab activities may help to promote expert-like student epistemologies and expectations about the nature of science in introductory courses.  Here, we explore the impact of open-ended activities more generally on students' epistemologies about and appreciation of experimental physics in lab courses both at \emph{and beyond} the introductory level.  To do this, we examine students' responses to the E-CLASS (Colorado Learning Attitudes about Science Survey for Experimental Physics) \cite{zwickl2012eclass}.  E-CLASS is a 30 item, Likert-style survey that targets students epistemologies and expectations about experimental physics, as well as student affect and confidence when doing physics experiments.  The E-CLASS presents students with a set of prompts (e.g., ``Calculating uncertainties helps me understand my results better.'') and asks them to rate their level of agreement both from their personal perspective when doing experiment in class and that of a hypothetical experimental physicist.  The E-CLASS was developed in conjunction with efforts to transform the upper-division laboratory courses at the University of Colorado Boulder (CU) \cite{zwickl2013adlab}.  The assessment was intended to be used in both introductory and advanced lab courses and, thus, includes items targeting a wide range of learning goals.  E-CLASS was validated through student interviews and expert review \cite{zwickl2014eclass}, and was tested for statistical validity and reliability using responses from students at multiple institutions and at multiple course levels \cite{wilcox2016eclass}.  This work is part of ongoing analysis of a growing, national data set of student responses to the E-CLASS.  

In this paper, we first describe the data sources (Sec.\ \ref{sec:data}) and analysis methods (Sec.\ \ref{sec:analysis}) used in this study.  We then present our findings with respect to whether the inclusion of open-ended activities was accompanied by improvements in students' postinstruction E-CLASS scores and examine how this varied for different course levels (Sec.\ \ref{sec:rawResults}).  In addition to examining raw postinstruction E-CLASS scores, we also determine whether the trends in postinstruction scores for different laboratory activities persisted after controlling for other factors such as preinstruction scores, course level, major, and gender (Sec.\ \ref{sec:ancovaResults}).  To investigate the relative effectiveness of different types of open-ended activities, we compare scores from students in courses using shorter-scale open-ended activities with those using longer-term, multi-week projects (Sec.\ \ref{sec:projectsResults}).  Finally, we end with a discussion of limitations of the study and future work (Sec.\ \ref{sec:discussion}).

\section{\label{sec:methods}Methods}

In this section, we present the data sources, student and institution demographics, and analysis methods used for this study.  

\subsection{\label{sec:data}Data sources}

This study is part of ongoing analysis of data collected using the E-CLASS centralized administration system \cite{wilcox2016admin} as part of a broader investigation of students' epistemologies in the context of physics lab courses (e.g., \cite{wilcox2016gender, wilcox2016pedagogy}).  The data set used here includes 3 semesters of E-CLASS responses collected between 01/2015 and 05/2016.  Students completed the E-CLASS online twice during the course, typically during the first and last weeks of class.  In addition to student responses to the E-CLASS prompts, the postinstruction version of the assessment also collected information on student demographics such as student major and gender.  

Only students with matched pre- and postinstruction E-CLASS responses were included in the analysis.  Pre to post matching was done based on student ID number or, when ID matching failed, by first and last name.  The E-CLASS also includes a filtering questions to eliminate responses from students who did not read the item prompts; any student who responded incorrectly to this filtering question was also dropped from the analysis (for more information see Ref.\ \cite{wilcox2016eclass}).  The final data set included $N=4915$ matched responses from 108 distinct courses at 67 institutions.  Based on estimates of the total enrollment provided by the instructors at the beginning of the course, this represents a matched response rate of roughly 40\%.  This response rate is only an approximation of the true response rate as enrollment may have fluctuated over the course of the semester.  The institutions in the data set spanned a range of institution types including 2-year ($N=3$) and 4-year colleges ($N=35$), as well as masters ($N=8$) and Ph.D. granting universities ($N=21$).  Several of these institutions used the E-CLASS in the same course during multiple semesters, thus the full data set includes student responses from 147 separate instances of the E-CLASS.  These courses also span multiple levels including first-year (FY) introductory courses and beyond-first-year (BFY) courses (Table \ref{tab:courseTypes}).

\begin{table}
\caption{Number of first-year and beyond-first-year courses in the matched data set.  The number of students in the beyond-first-year courses is smaller in part because of the smaller class sizes typical of more advanced physics labs.  The number of separate instances of the E-CLASS accounts for courses that administered E-CLASS more than once in the 3 semesters of data collection.   }\label{tab:courseTypes}
\begin{ruledtabular}
   \begin{tabular}{ l r r r }
       & Distinct & Separate & \hspace{1mm} Number of \\
       &\hspace{1mm} courses &\hspace{1mm} instances & Students  \\
     \hline
     First-year & 49 & 71 & 4083  \\
     Beyond-first-year \hspace{1mm}& 59 & 76 & 832  \\
   \end{tabular}
\end{ruledtabular}
\end{table}  

\begin{table*}
\caption{Demographic breakdown of the full data set, as well as the subset of courses with open-ended activities and those with only traditional guided lab activities.  Number of courses refers to the number of distinct courses, and percentages represent the percentage of students rather than the percentage of courses.  For Major and Gender demographics, the totals may not sum to 100\% as some students did not complete these questions or selected `Other' as their gender.     }\label{tab:dems}
\begin{ruledtabular}
   \begin{tabular}{ l c c c c c c c c }
       & \multicolumn{2}{c}{N} & \multicolumn{2}{c}{Course Level} & \multicolumn{2}{c}{Gender} & \multicolumn{2}{c}{Major} \\
       & Courses & Students & FY & BFY & Women & Men & Physics & Non-physics  \\
     \hline
     All Courses & 147 & 4915 & 83\% & 17\% & 40\% & 57\% & 21\% & 78\%  \\
     Open-ended & 84 & 1149 & 52\% & 48\%  & 36\% & 62\% & 49\% & 50\%  \\
     Guided only & 63 & 3766 & 93\% & 7\% & 42\% & 56\% & 12\% & 87\%  \\
   \end{tabular}
\end{ruledtabular}
\end{table*}  

\begin{figure}[b]
\includegraphics[width=\linewidth]{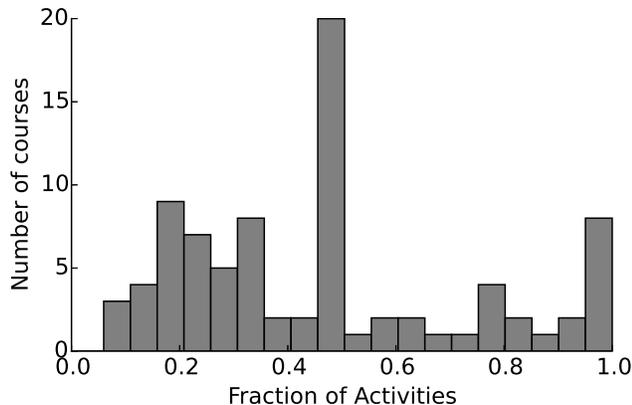}
\caption{Histogram of the fraction of weeks spent on open-ended activities for the $N=84$ courses that reported including one or more weeks of open-ended activities.}\label{fig:activitiesHist}
\end{figure}

In order to use E-CLASS through its centralized administration system \cite{wilcox2016admin}, instructors complete a Course Information Survey (CIS) in which they report basic information about their course and institution.  The CIS collects both logistical information (e.g., estimated enrollment, course start and end dates, course syllabi, etc.) as well as information on course activities and the instructors' use of various pedagogical techniques.  On the CIS, instructors were asked to report how many weeks of the semester were spent on ``all guided lab activities'' and how many weeks were spent on ``all open-ended activities or projects.''  The terms ``guided lab'' and ``open-ended'' activities were not operationalized in the CIS; thus, instructors responses are self-reported and self-classified.  While the relative fraction of the course spent on open-ended activities varied significantly (see Fig.\ \ref{fig:activitiesHist}), 84 courses reported having at least one week of open-ended activities.  To preserve statistical power, the remainder of this analysis will treat courses dichotomously as either having open-ended activities (regardless of the fraction of the course those activities represent) or having only traditional guided lab activities.  

The the demographic breakdown of the matched E-CLASS data set in terms of course level, major, and gender is given in Table \ref{tab:dems}.  Table \ref{tab:dems} also compares the breakdown of these data across courses that were classified as including open-ended activities and those with only traditional guided labs.  A larger fraction of the courses that included some open-ended activities were also BFY courses.  This trend may be driven in part by the smaller class sizes characteristic of BFY courses, as open-ended activities often require lower student-to-teacher ratios in order to provide sufficient instructor support to the students.  Table \ref{tab:dems} does not include racial demographic data because these data were collected only in the final two semesters of data collection. Examination of E-CLASS scores with regards to racial dynamics will be the subject of a future publication. In addition to gender data, the postinstruction E-CLASS also asked students for their primary major.  While students were offered 15 distinct major options, we focus here on students' major as the dichotomous distinction between `physics' or `non-physics' majors, where physics includes both engineering physics and physics majors, and non-physics includes all other majors (including other science majors, non-science majors, and students who are open-option or undeclared).  The variations in the prior and ongoing experiences
of students in different non-physics majors are likely significant; however, clearly characterizing the nature of these differences given the large number of courses and institutions in the data set is not possible. Given this, and the physics focus of the E-CLASS, we have chosen to focus or analysis of student major specifically on the difference between physics and non-physics majors.

\subsection{\label{sec:analysis}Analysis}

For the purposes of scoring the E-CLASS, we collapsed students' responses to each 5-point Likert item into a standardized, 3-point scale in which the responses `(dis)agree' and `strongly (dis)agree' were collapsed into a single category.  Students' responses to individual items were given a numerical score based on their consistency with the accepted, expert-like response \cite{zwickl2012eclass}: $+1$ for favorable (i.e., consistent with experts); $+0$ for neutral; and $-1$ for unfavorable (i.e., inconsistent with experts).  A student's overall E-CLASS score was then given by the sum of their scores on each of the 30 items resulting in a possible range of scores of $[-30,30]$.  For more information on the scoring of the E-CLASS see Ref. \cite{wilcox2016eclass}.  In previous work, we have cautioned instructors using the E-CLASS against focusing exclusively on the overall score when interpreting their results \cite{wilcox2016eclass}. The E-CLASS targets a range of learning goals some of which may not be relevant to a specific course, and we encourage instructors to focus also on the individual items most relevant to their learning goals.  For this reason, we provide also a breakdown of students' scores by item. However, the overall score is still useful in that it provides a continuous variable that offers a wholistic view of students' performance on the E-CLASS that can be used to quantitatively examine how that performance varies across subpopulations of students.  As the distribution of scores on the E-CLASS is typically skewed towards positive scores \cite{wilcox2016eclass,wilcox2016gender}, the following sections report statistical significance based on the non-parametric Mann-Whitney U test \cite{mann1947mwu} unless otherwise stated.  Where differences between means are statistically significant, we also report Cohen's $d$ \cite{cohen1988d} as a measure of effect size and practical significance \cite{rodriguez2012equity}. 

\begin{figure*}
\includegraphics[width=\linewidth]{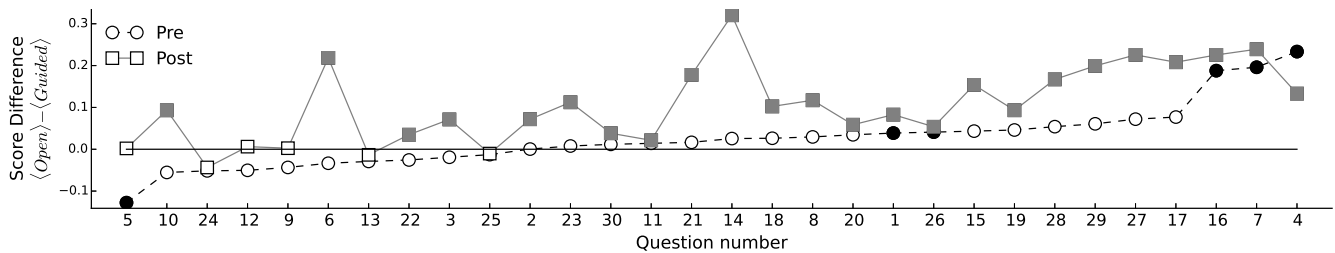}
\caption{Sorted plot of the difference between the average scores (points) of students in courses with open-ended activities and those with only traditional guided lab activities for each item of the E-CLASS. Zero difference is marked by the solid horizontal line.  Filled markers indicate points for which the difference between the distributions of students in open-ended and guided lab only courses was statistically significant (Mann-Whitney U \cite{mann1947mwu} and Holm-Bonferroni \cite{holm1979hb} corrected $p<0.05$).  See Ref.\ \cite{ECLASSwebsite} for full list of item prompts. }\label{fig:byItem}
\end{figure*}

Table \ref{tab:dems} highlights demographic differences between courses using open-ended activities and those using only traditional guided labs.  As has been observed previously in student responses to the E-CLASS \cite{wilcox2016gender, wilcox2016pedagogy}, these demographic differences may confound comparisons of students E-CLASS scores in these two types of courses.  To account for this, we utilize an analysis of covariance (ANCOVA) \cite{wildt1978ancova} in addition to examining students' raw pre- and postinstruction E-CLASS scores.  ANCOVA is a statistical method for comparing the difference between population means after adjusting them to account for the variance associated with other variables.  In order for the results of an ANCOVA to be valid, the data must meet several assumptions.  The assumptions of an ANCOVA are discussed in detail in Refs. \cite{wildt1978ancova, day2016gender}; tests of the E-CLASS matched data showed that they satisfied these assumptions, with one exception.  In our data, the covariate (i.e., preinstruction score) is not independent of the other variables (i.e., gender, major, and course level).  Shared variance between the covariate and independent variables is to be expected in any observational study in which randomized assignment to experimental groups was not done or not possible \cite{miller2001ancova}.  Violation of the assumption of covariate independence implies that our results should be interpreted as a lower bound on the relationship between each independent variable and postinstruction E-CLASS score.

\section{\label{sec:results}Results}

This section presents findings with respect to whether the inclusion of open-ended activities was accompanied by improvements in students' postinstruction E-CLASS responses using raw scores and an ANCOVA.

\subsection{\label{sec:rawResults}Open-ended vs. traditional guided lab activities}

To explore general trends in the aggregate data, we first examine differences in raw pre- and postinstruction E-CLASS scores for students in courses using open-ended activities and those using only traditional guided labs (Table \ref{tab:overallAgg}).  Table \ref{tab:overallAgg} shows that students in courses using open-ended activities scored significantly higher than students in courses using only traditional guided labs both pre- and postinstruction ($p\ll0.01$).  While the difference is statistically significant both before and after instruction, the magnitude of this effect was larger for the postinstruction scores.  Moreover, students in courses using open-ended activities showed a small ($d=0.08$), but statistically significant, positive shift ($p\ll0.01$) before and after instruction, while students in courses using only traditional guided labs showed a small ($d=-0.2$) but statistically significant negative shift ($p\ll0.01$).  

\begin{table}
\caption{Overall E-CLASS scores (points) for students in courses using open-ended activities and those using only traditional guided labs in the full, aggregate data set ($N=4915$) on both the pre- and post-tests.  Significance indicates the statistical significance of the difference between students' scores in the open-ended and guided lab only courses.  }\label{tab:overallAgg}
\begin{ruledtabular}
   \begin{tabular}{ l r r r r}
       &Open-ended& Guided& \hspace{2mm} Significance & \hspace{2mm} Effect Size\\
      \hline
      N & 1149 & 3766 & - & - \\
      Pre & 16.9 & 16.2 & $p\ll0.01$ & $d=0.1$ \\
      Post & 17.5 & 14.4 & $p\ll0.01$ & $d=0.4$ \\
   \end{tabular}
\end{ruledtabular}
\end{table}

We can also examine the difference between courses using open-ended activities and courses using only traditional guided labs on an item-by-item scale.  Fig.\ \ref{fig:byItem} shows the difference between the average scores of students in these two types of courses for each of the 30 items on the pre- and post-instruction E-CLASS.  Consistent with the small difference in the overall preinstruction score (Table \ref{tab:overallAgg}), only 6 items show a statistically significant difference in the distribution of scores between open-ended and guided lab activities.  For 5 of these 6 questions, students in open-ended courses scored higher, and the magnitude of the difference was small in all cases ($d<0.3$).  Alternatively, 24 of the 30 items showed statistically significant differences after instruction with students in courses using open-ended activities outscoring students in courses using only traditional guided labs in all cases.  The magnitude of the difference was moderate ($d=0.4$) for only one item (Item 14: ``When doing an experiment I usually think up my own questions to investigate''), and small ($d<0.3$) for the remaining 23 items.  

While comparisons of E-CLASS scores both overall and by-item in the full, aggregate data set preliminarily suggest that open-ended activities have a positive impact on students' performance, the statistically significant difference in preinstruction scores between students suggests that these two types of courses may have served different student populations.  This conclusion is supported by the demographic data presented in Table \ref{tab:dems}, which, for example, shows that courses using open-ended activities are considerably more likely to also be BFY courses.  Previous work has shown that students in BFY courses consistently score higher on E-CLASS than students in FY courses \cite{wilcox2016eclass}, potentially accounting for some of the difference in preinstruction scores between open-ended and guided lab courses.  

To determine whether the relationship between the type of activities used and postinstruction E-CLASS score varies across course levels, we examined the overall E-CLASS scores for students in FY and BFY courses separately.  The trend of higher postinstruction scores in courses using open-ended activities persisted for both FY and BFY students (see Table \ref{tab:courseLevelAvg}); however, the pattern of shifts varies between the two types of courses.  In FY courses, students in courses using open-ended activities did not show statistically significant shift from pre- to postinstruction ($p=0.4$), while students in courses using only guided activities only showed a small ($d=-0.2$), but statistically significant, negative shift  ($p\ll0.01$).  Alternatively, in BFY courses, courses using open-ended activities showed a small ($d=0.2$) but statistically significant positive shift from pre- to postinstruction ($p\ll0.01$), while courses using only guided activities showed a small ($d=-0.1$) negative shift ($p=0.03$).  

\begin{table}[b]
\caption{Overall E-CLASS scores (points) for students in courses using open-ended activities and those using only traditional guided labs in the FY and BFY student populations separately.  Significance indicates the statistical significance of the difference between students' scores in the open-ended and guided lab only courses.  } \label{tab:courseLevelAvg}
\begin{ruledtabular}
   \begin{tabular}{ l l r r r r}
      Level & &Open-ended& Guided& Significance & Effect Size\\
      \hline
      &N & 592 & 3487 & - & - \\
      FY & Pre & 15.4 & 16.0 & $p=0.02$ & $d=-0.09$ \\
      &Post & 15.5 & 14.1 & $p\ll0.01$ & $d=0.2$ \\
      \hline
      &N & 553 & 279 & - & - \\
      BFY &Pre & 18.6 & 18.1 & $p=0.5$ &  \\
      &Post & 19.7 & 17.2 & $p\ll0.01$ & $d=0.3$ \\
   \end{tabular}
\end{ruledtabular}
\end{table}

Table \ref{tab:courseLevelAvg} summarizes the effect of one potential confounding variable (i.e., course level) on the analysis of the the relationship between open-ended activities and E-CLASS scores; however, there may be other variables to take into consideration.  For example, student responses to the E-CLASS have been shown to vary based on students' major \cite{wilcox2016eclass}.  Additionally, prior research suggests that some transformed instructional approaches may have a differentially positive impact on the E-CLASS scores of women \cite{wilcox2016pedagogy}, suggesting that student gender may also be a significant factor.  Sec.\ \ref{sec:ancovaResults} explores these dynamics using an analysis of covariance.

\subsection{\label{sec:ancovaResults}Analysis of covariance}

To more clearly explore the relationship between open-ended activities and post-instruction E-CLASS scores independent from other factors, we used an ANCOVA.  ANCOVA is a statistical method for comparing the difference between population means while adjusting them to account for the variance associated with other variables.  In this case, we want to determine if the difference between the E-CLASS scores of students in courses using different types of lab activities (open-ended vs.\ guided only) remains statistically significant after accounting for differences in preinstruction scores, as well as student major and gender.  Only students for whom we have matched E-CLASS scores along with both major and gender data were included in the ANCOVA ($N=4759$).  

We performed a 5-way ANCOVA to compare post-instruction E-CLASS means for courses using open-ended and guided lab activities while controlling for the three categorical variables: course level, student major, and student gender, as well as preinstruction score as a covariate.  To determine how these variables might also be related to one another, we initially included all possible interaction terms.  None of the interaction terms were statistically significant predictors of postinstruction scores.  This result should be interpreted as evidence that the impact of open-ended activities did not vary significantly depending on the other variables.  For example, these data do not suggest that open-ended activities had a more positive impact on women than men.  As the interaction terms did not contribute significantly, they were removed from the model.  

The results of the 5-way ANCOVA (without interaction terms) are summarized in Table \ref{tab:ancova}.  All four categorical variables (gender, major, course level, and type of activities) were statistically significant predictors of postinstruction E-CLASS score (F-test, $p<0.01$).  Type of activities (open-ended vs.\ guided only) accounted for the largest difference in adjusted postinstruction means with students in courses using open-ended activities scoring higher.  Thus, when adjusting for the variance associated with preinstruction score, course level, major, and gender, students in courses using open-ended activities demonstrate more expert-like E-CLASS responses than those in courses using only traditional guided labs.  

\begin{table}
\caption{Comparison of postinstruction means as adjusted by the 5-way ANCOVAs for each categorical variable.  A difference between group means is indicated only when that difference was statistically significant.  Here, $\langle O\rangle$ is the predicted postinstruction mean for students in courses using open-ended activities and similarly for students in courses using only guided activities $\langle G\rangle$, physics students $\langle P\rangle$, non-physics students $\langle NP\rangle$, men $\langle M\rangle$, women $\langle W \rangle$, BFY students $\langle BFY\rangle$, and FY students $\langle FY \rangle$.  Variables are listed in descending order by size of the difference in adjusted postinstruction means between groups.   } \label{tab:ancova}
\begin{ruledtabular}
   \begin{tabular}{ l c }
       Catagorical Variable & Postinstruction mean comparison\\
      \hline      
      Activities & $\langle O\rangle>\langle G\rangle$ \\
      Major & $\langle P\rangle>\langle NP\rangle$ \\      
      Course level \hspace{4mm} & $\langle BFY\rangle>\langle FY\rangle$ \\
      Gender & $\langle M\rangle>\langle W\rangle$ \\
   \end{tabular}
\end{ruledtabular}
\end{table}


\subsection{\label{sec:projectsResults}Open-ended activities vs. multi-week projects}

The results presented in the previous sections support the idea that the use of open-ended activities in undergraduate lab courses improved students epistemologies, expectation, and confidence with respect to the nature of experimental physics.  However, courses in the data set were self-classified by instructors as including open-ended activities; thus, there is likely significant variation in the types of open-ended activities represented in the data.  Moreover, it is likely that not all open-ended activities are equally effective at encouraging expert-like epistemologies and expectations.  For example, we argue that longer term, multi-week projects have the potential to provide some of the most authentic experimental physics activities that an undergraduate student might engage in outside of undergraduate research.  Because of this, we hypothesized that courses that included multi-week projects would result in higher E-CLASS scores than shorter, week-to-week open-ended activities.  

Whether a course included a multi-week project was not specifically asked on the CIS; however, the CIS did collect course syllabi, which generally include a description of the course activities and expectations and/or a grading breakdown showing the fraction of the grade from each of the activities in the course.  Courses were coded as having a project component if the syllabus listed a project in either the course description or grading breakdown.  In a few cases, we were not able to determine if the course included a project because the syllabus was unclear or missing.  In our data set, all of the 22 courses ($N=231$) that were identified as including a project component were BFY, and all were classified by the instructor as including open-ended activities.  To account for this, the comparison group was the 31 BFY courses ($N=306$) that included open-ended activities but whose syllabi clearly indicated they did not include a project component.  

\begin{table}[b]
\caption{Overall E-CLASS scores (points) for students in courses that included multi-week projects and those that did not.  Consistent with the courses that include projects, all courses represented here are BFY courses that are self-classified as including open-ended activities.  Significance indicates the statistical significance of the difference between students' scores in the project and non-project courses.  }\label{tab:projects}
\begin{ruledtabular}
   \begin{tabular}{ l r r r }
       & Project & Non-project & Significance \\
      \hline
      N & 231 & 306 & - \\
      Pre & 18.4 & 18.7 & $p=0.1$ \\
      Post & 19.8 & 19.6 & $p=0.5$ \\
   \end{tabular}
\end{ruledtabular}
\end{table}

Table \ref{tab:projects} shows the pre- and postinstruction scores for courses that include a project and those that do not.  There was no statistically significant difference in the average E-CLASS scores either before or after instruction for students in these two sets of courses.  This result indicates that, while courses with multi-week projects did score significantly higher than courses using only traditional guided labs, they did not result in a significant increase in E-CLASS scores above and beyond that of shorter-term open-ended activities.  However, this finding should not be interpreted as evidence that projects do not contribute significantly to lab courses in ways beyond that of general open-ended activities.  As discussed in Sec.\ \ref{sec:intro}, lab courses have a multiple different learning goals, and E-CLASS targets only a subset of these goals.  The inclusion of multi-week projects may have significant impact on these other learning goals (e.g., developing student ownership, or practical and managerial lab skills).  

\section{\label{sec:discussion}Summary and Conclusions}

We analyzed a large data set of student responses to the E-CLASS for evidence of the impact of open-ended laboratory activities on students' epistemologies, expectations, and confidence with respect to experimental physics.  We found that courses that included open-ended activities during one or more weeks of the laboratory had higher pre- and postinstruction E-CLASS scores as well as more favorable shifts relative to courses using only traditional guided lab activities.  This result was reinforced by an analysis of covariance, which showed that the type of activity used (open-ended vs. guided only) was a significant predictor of postinstruction E-CLASS score even after adjusting for the variance associated with preinstruction score, course level, student major, and student gender.  We also examined the effectiveness of multi-week projects relative to shorter-term open-ended activities and found no evidence that multi-week projects resulted in more expert-like E-CLASS responses than open-ended activities generally.  

Overall, our findings support the claim that the use of open-ended activities may have a positive impact on students epistemologies about the nature of experimental physics and their affect and confidence when performing physics experiments.  We also found that this positive impact does not require implementation of multi-week projects.  However, there are several limitations of this work.  While our data set is extensive, spanning a large number of institutions, courses, and student populations, it is not comprehensive. For example, there are only a few 2-year colleges in our data. Additionally, we focused here on a specific subset of potential variables that might confound the comparison of courses using different types of activities (i.e., major, course level, and preinstruction scores). These variables were selected based on the findings of previous work \cite{wilcox2016eclass,wilcox2016gender,wilcox2016pedagogy}, which suggested they were important factors in predicting postinstruction E-CLASS scores. However, there are other factors that might also correlate with they type of activities used by the instructor, including class size, instructor familiarity with PER, and student-to-teacher ratio.  Similarly, the instructors for the courses in our data set generally chose to use E-CLASS without external pressure, and thus these courses are not randomly selected. Additionally, to preserve statistical power, all courses using any open-ended activities were aggregated together as a single group. Thus, while it may be that having a greater fraction of the course dedicated to open-ended activities would be more effective at promoting expert-like epistemologies and expectations, the current data set cannot address this dynamic. As data collection with the E-CLASS centralized administration system continues, examination of the impact of more frequent open-ended activities may become possible. 

Additionally, the purely quantitative analysis reported here cannot speak to how open-ended activities may have improved students' epistemologies and expectations relative to traditional approaches. We hypothesized that open-ended activities may provide greater opportunities for the students to engage authentically in the process of experimental physics; however, clearly determining the mechanism underlying the findings reported here will likely require additional research with a significant qualitative component.  Future work could also include more fine-grained investigation of the relative effectiveness of different types of open-ended activities.  Here, we examined the relative effectiveness of multi-week projects relative to other open-ended activities; however, there are many other types of open-ended activities that may be more or less effective at encouraging expert-like E-CLASS responses.  Investigations of this type would require the creation of a robust and valid classification scheme for different open-ended activities.  The development and implementation of such a scheme might require collection of course artifacts (e.g., lab manuals) or in-class observations.


\begin{acknowledgments}
This work was funded by the NSF-IUSE Grant No. DUE-1432204 and NSF Grant No. PHY-1125844.  Particular thanks to the members of PER@C for all their help and feedback.  
\end{acknowledgments}

\bibliography{master-refs-ECLASS-5_16}

\begin{thebibliography}{28}%
\makeatletter
\providecommand \@ifxundefined [1]{%
 \@ifx{#1\undefined}
}%
\providecommand \@ifnum [1]{%
 \ifnum #1\expandafter \@firstoftwo
 \else \expandafter \@secondoftwo
 \fi
}%
\providecommand \@ifx [1]{%
 \ifx #1\expandafter \@firstoftwo
 \else \expandafter \@secondoftwo
 \fi
}%
\providecommand \natexlab [1]{#1}%
\providecommand \enquote  [1]{``#1''}%
\providecommand \bibnamefont  [1]{#1}%
\providecommand \bibfnamefont [1]{#1}%
\providecommand \citenamefont [1]{#1}%
\providecommand \href@noop [0]{\@secondoftwo}%
\providecommand \href [0]{\begingroup \@sanitize@url \@href}%
\providecommand \@href[1]{\@@startlink{#1}\@@href}%
\providecommand \@@href[1]{\endgroup#1\@@endlink}%
\providecommand \@sanitize@url [0]{\catcode `\\12\catcode `\$12\catcode
  `\&12\catcode `\#12\catcode `\^12\catcode `\_12\catcode `\%12\relax}%
\providecommand \@@startlink[1]{}%
\providecommand \@@endlink[0]{}%
\providecommand \url  [0]{\begingroup\@sanitize@url \@url }%
\providecommand \@url [1]{\endgroup\@href {#1}{\urlprefix }}%
\providecommand \urlprefix  [0]{URL }%
\providecommand \Eprint [0]{\href }%
\providecommand \doibase [0]{http://dx.doi.org/}%
\providecommand \selectlanguage [0]{\@gobble}%
\providecommand \bibinfo  [0]{\@secondoftwo}%
\providecommand \bibfield  [0]{\@secondoftwo}%
\providecommand \translation [1]{[#1]}%
\providecommand \BibitemOpen [0]{}%
\providecommand \bibitemStop [0]{}%
\providecommand \bibitemNoStop [0]{.\EOS\space}%
\providecommand \EOS [0]{\spacefactor3000\relax}%
\providecommand \BibitemShut  [1]{\csname bibitem#1\endcsname}%
\let\auto@bib@innerbib\@empty
\bibitem [{\citenamefont {Trumper}(2003)}]{trumper2003labs}%
  \BibitemOpen
  \bibfield  {author} {\bibinfo {author} {\bibfnamefont {Ricardo}\ \bibnamefont
  {Trumper}},\ }\bibfield  {title} {\enquote {\bibinfo {title} {The physics
  laboratory -- a historical overview and future perspectives},}\ }\href
  {\doibase 10.1023/A:1025692409001} {\bibfield  {journal} {\bibinfo  {journal}
  {Science {\&} Education}\ }\textbf {\bibinfo {volume} {12}},\ \bibinfo
  {pages} {645--670} (\bibinfo {year} {2003})}\BibitemShut {NoStop}%
\bibitem [{\citenamefont {Zwickl}\ \emph {et~al.}(2013)\citenamefont {Zwickl},
  \citenamefont {Finkelstein},\ and\ \citenamefont
  {Lewandowski}}]{zwickl2013adlab}%
  \BibitemOpen
  \bibfield  {author} {\bibinfo {author} {\bibfnamefont {Benjamin~M}\
  \bibnamefont {Zwickl}}, \bibinfo {author} {\bibfnamefont {Noah}\ \bibnamefont
  {Finkelstein}}, \ and\ \bibinfo {author} {\bibfnamefont {HJ}~\bibnamefont
  {Lewandowski}},\ }\bibfield  {title} {\enquote {\bibinfo {title} {The process
  of transforming an advanced lab course: Goals, curriculum, and
  assessments},}\ }\href@noop {} {\bibfield  {journal} {\bibinfo  {journal}
  {American Journal of Physics}\ }\textbf {\bibinfo {volume} {81}},\ \bibinfo
  {pages} {63--70} (\bibinfo {year} {2013})}\BibitemShut {NoStop}%
\bibitem [{\citenamefont {{AAPT Committee on
  Laboratories}}(2015)}]{AAPT2015guidelines}%
  \BibitemOpen
  \bibfield  {author} {\bibinfo {author} {\bibnamefont {{AAPT Committee on
  Laboratories}}},\ }\href@noop {} {\enquote {\bibinfo {title} {{AAPT
  Recommendations for the Undergraduate Physics Laboratory Curriculum}},}\ }
  (\bibinfo {year} {2015})\BibitemShut {NoStop}%
\bibitem [{\citenamefont {Wieman}(2015)}]{wieman2015labs}%
  \BibitemOpen
  \bibfield  {author} {\bibinfo {author} {\bibfnamefont {Carl}\ \bibnamefont
  {Wieman}},\ }\bibfield  {title} {\enquote {\bibinfo {title} {Comparative
  cognitive task analyses of experimental science and instructional laboratory
  courses},}\ }\href@noop {} {\bibfield  {journal} {\bibinfo  {journal} {The
  Physics Teacher}\ }\textbf {\bibinfo {volume} {53}} (\bibinfo {year}
  {2015})}\BibitemShut {NoStop}%
\bibitem [{\citenamefont {{National Research Council}}(2013)}]{nap2013nrc}%
  \BibitemOpen
  \bibfield  {author} {\bibinfo {author} {\bibnamefont {{National Research
  Council}}},\ }\href@noop {} {\emph {\bibinfo {title} {Adapting to a Changing
  World--Challenges and Opportunities in Undergraduate Physics Education}}}\
  (\bibinfo  {publisher} {The National Academies Press},\ \bibinfo {year}
  {2013})\BibitemShut {NoStop}%
\bibitem [{\citenamefont {Etkina}\ and\ \citenamefont
  {Heuvelen}(2001)}]{etkina2001isle}%
  \BibitemOpen
  \bibfield  {author} {\bibinfo {author} {\bibfnamefont {Eugenia}\ \bibnamefont
  {Etkina}}\ and\ \bibinfo {author} {\bibfnamefont {Alan~Van}\ \bibnamefont
  {Heuvelen}},\ }\bibfield  {title} {\enquote {\bibinfo {title} {Investigative
  science learning environment: Using the processes of science and cognitive
  strategies to learn physics},}\ }in\ \href@noop {} {\emph {\bibinfo
  {booktitle} {Physics Education Research Conference 2001}}},\ \bibinfo {series
  and number} {PER Conference}\ (\bibinfo {address} {Rochester, New York},\
  \bibinfo {year} {2001})\BibitemShut {NoStop}%
\bibitem [{\citenamefont {Wells}\ \emph {et~al.}(1995)\citenamefont {Wells},
  \citenamefont {Hestenes},\ and\ \citenamefont
  {Swackhamer}}]{wells1995modeling}%
  \BibitemOpen
  \bibfield  {author} {\bibinfo {author} {\bibfnamefont {Malcolm}\ \bibnamefont
  {Wells}}, \bibinfo {author} {\bibfnamefont {David}\ \bibnamefont {Hestenes}},
  \ and\ \bibinfo {author} {\bibfnamefont {Gregg}\ \bibnamefont {Swackhamer}},\
  }\bibfield  {title} {\enquote {\bibinfo {title} {A modeling method for high
  school physics instruction},}\ }\href@noop {} {\bibfield  {journal} {\bibinfo
   {journal} {Am. J. Phys.}\ }\textbf {\bibinfo {volume} {63}},\ \bibinfo
  {pages} {606--619} (\bibinfo {year} {1995})}\BibitemShut {NoStop}%
\bibitem [{\citenamefont {Wilson}(1994)}]{wilson1994studio}%
  \BibitemOpen
  \bibfield  {author} {\bibinfo {author} {\bibfnamefont {Jack~M.}\ \bibnamefont
  {Wilson}},\ }\bibfield  {title} {\enquote {\bibinfo {title} {The cuple
  physics studio},}\ }\href {\doibase http://dx.doi.org/10.1119/1.2344100}
  {\bibfield  {journal} {\bibinfo  {journal} {The Physics Teacher}\ }\textbf
  {\bibinfo {volume} {32}},\ \bibinfo {pages} {518--523} (\bibinfo {year}
  {1994})}\BibitemShut {NoStop}%
\bibitem [{\citenamefont {Beichner}\ \emph {et~al.}(2000)\citenamefont
  {Beichner}, \citenamefont {Saul}, \citenamefont {Allain}, \citenamefont
  {Deardorff},\ and\ \citenamefont {Abbott}}]{beichner2000scaleup}%
  \BibitemOpen
  \bibfield  {author} {\bibinfo {author} {\bibfnamefont {Robert~J}\
  \bibnamefont {Beichner}}, \bibinfo {author} {\bibfnamefont {Jeffery~M}\
  \bibnamefont {Saul}}, \bibinfo {author} {\bibfnamefont {Rhett~J}\
  \bibnamefont {Allain}}, \bibinfo {author} {\bibfnamefont {Duane~L}\
  \bibnamefont {Deardorff}}, \ and\ \bibinfo {author} {\bibfnamefont {David~S}\
  \bibnamefont {Abbott}},\ }\bibfield  {title} {\enquote {\bibinfo {title}
  {Introduction to scale-up: Student-centered activities for large enrollment
  university physics.}}\ }\href@noop {} {\  (\bibinfo {year}
  {2000})}\BibitemShut {NoStop}%
\bibitem [{\citenamefont {Beichner}\ \emph {et~al.}(2007)\citenamefont
  {Beichner}, \citenamefont {Saul}, \citenamefont {Abbott}, \citenamefont
  {Morse}, \citenamefont {Deardorff}, \citenamefont {Allain}, \citenamefont
  {Bonham}, \citenamefont {Dancy},\ and\ \citenamefont
  {Risley}}]{beichner2007scaleup}%
  \BibitemOpen
  \bibfield  {author} {\bibinfo {author} {\bibfnamefont {Robert~J}\
  \bibnamefont {Beichner}}, \bibinfo {author} {\bibfnamefont {Jeffery~M}\
  \bibnamefont {Saul}}, \bibinfo {author} {\bibfnamefont {David~S}\
  \bibnamefont {Abbott}}, \bibinfo {author} {\bibfnamefont {Jeanne~J}\
  \bibnamefont {Morse}}, \bibinfo {author} {\bibfnamefont {Duane}\ \bibnamefont
  {Deardorff}}, \bibinfo {author} {\bibfnamefont {Rhett~J}\ \bibnamefont
  {Allain}}, \bibinfo {author} {\bibfnamefont {Scott~W}\ \bibnamefont
  {Bonham}}, \bibinfo {author} {\bibfnamefont {Melissa~H}\ \bibnamefont
  {Dancy}}, \ and\ \bibinfo {author} {\bibfnamefont {John~S}\ \bibnamefont
  {Risley}},\ }\bibfield  {title} {\enquote {\bibinfo {title} {The
  student-centered activities for large enrollment undergraduate programs
  (scale-up) project},}\ }\href@noop {} {\bibfield  {journal} {\bibinfo
  {journal} {Research-based reform of university physics}\ }\textbf {\bibinfo
  {volume} {1}},\ \bibinfo {pages} {2--39} (\bibinfo {year}
  {2007})}\BibitemShut {NoStop}%
\bibitem [{\citenamefont {Adams}\ \emph {et~al.}(2006)\citenamefont {Adams},
  \citenamefont {Perkins}, \citenamefont {Podolefsky}, \citenamefont {Dubson},
  \citenamefont {Finkelstein},\ and\ \citenamefont {Wieman}}]{adams2006class}%
  \BibitemOpen
  \bibfield  {author} {\bibinfo {author} {\bibfnamefont {Wendy~K}\ \bibnamefont
  {Adams}}, \bibinfo {author} {\bibfnamefont {Katherine~K}\ \bibnamefont
  {Perkins}}, \bibinfo {author} {\bibfnamefont {Noah~S}\ \bibnamefont
  {Podolefsky}}, \bibinfo {author} {\bibfnamefont {Michael}\ \bibnamefont
  {Dubson}}, \bibinfo {author} {\bibfnamefont {Noah~D}\ \bibnamefont
  {Finkelstein}}, \ and\ \bibinfo {author} {\bibfnamefont {Carl~E}\
  \bibnamefont {Wieman}},\ }\bibfield  {title} {\enquote {\bibinfo {title} {New
  instrument for measuring student beliefs about physics and learning physics:
  The colorado learning attitudes about science survey},}\ }\href@noop {}
  {\bibfield  {journal} {\bibinfo  {journal} {Physical Review Special
  Topics-Physics Education Research}\ }\textbf {\bibinfo {volume} {2}},\
  \bibinfo {pages} {010101} (\bibinfo {year} {2006})}\BibitemShut {NoStop}%
\bibitem [{\citenamefont {Redish}\ \emph {et~al.}(1998)\citenamefont {Redish},
  \citenamefont {Saul},\ and\ \citenamefont {Steinberg}}]{redish1998mpex}%
  \BibitemOpen
  \bibfield  {author} {\bibinfo {author} {\bibfnamefont {Edward~F}\
  \bibnamefont {Redish}}, \bibinfo {author} {\bibfnamefont {Jeffery~M}\
  \bibnamefont {Saul}}, \ and\ \bibinfo {author} {\bibfnamefont {Richard~N}\
  \bibnamefont {Steinberg}},\ }\bibfield  {title} {\enquote {\bibinfo {title}
  {Student expectations in introductory physics},}\ }\href@noop {} {\bibfield
  {journal} {\bibinfo  {journal} {American Journal of Physics}\ }\textbf
  {\bibinfo {volume} {66}},\ \bibinfo {pages} {212--224} (\bibinfo {year}
  {1998})}\BibitemShut {NoStop}%
\bibitem [{\citenamefont {Brewe}\ \emph {et~al.}(2009)\citenamefont {Brewe},
  \citenamefont {Kramer},\ and\ \citenamefont {O’Brien}}]{brewe2009modeling}%
  \BibitemOpen
  \bibfield  {author} {\bibinfo {author} {\bibfnamefont {Eric}\ \bibnamefont
  {Brewe}}, \bibinfo {author} {\bibfnamefont {Laird}\ \bibnamefont {Kramer}}, \
  and\ \bibinfo {author} {\bibfnamefont {George}\ \bibnamefont {O’Brien}},\
  }\bibfield  {title} {\enquote {\bibinfo {title} {Modeling instruction:
  Positive attitudinal shifts in introductory physics measured with class},}\
  }\href@noop {} {\bibfield  {journal} {\bibinfo  {journal} {Physical Review
  Special Topics-Physics Education Research}\ }\textbf {\bibinfo {volume}
  {5}},\ \bibinfo {pages} {013102} (\bibinfo {year} {2009})}\BibitemShut
  {NoStop}%
\bibitem [{\citenamefont {Kohl}\ and\ \citenamefont
  {Vincent~Kuo}(2012)}]{kohl2012studio}%
  \BibitemOpen
  \bibfield  {author} {\bibinfo {author} {\bibfnamefont {Patrick~B.}\
  \bibnamefont {Kohl}}\ and\ \bibinfo {author} {\bibfnamefont {H.}~\bibnamefont
  {Vincent~Kuo}},\ }\bibfield  {title} {\enquote {\bibinfo {title} {Chronicling
  a successful secondary implementation of studio physics},}\ }\href@noop {}
  {\bibfield  {journal} {\bibinfo  {journal} {American Journal of Physics}\
  }\textbf {\bibinfo {volume} {80}},\ \bibinfo {pages} {832--839} (\bibinfo
  {year} {2012})}\BibitemShut {NoStop}%
\bibitem [{\citenamefont {Zwickl}\ \emph {et~al.}(2012)\citenamefont {Zwickl},
  \citenamefont {Finkelstein},\ and\ \citenamefont
  {Lewandowski}}]{zwickl2012eclass}%
  \BibitemOpen
  \bibfield  {author} {\bibinfo {author} {\bibfnamefont {Benjamin}\
  \bibnamefont {Zwickl}}, \bibinfo {author} {\bibfnamefont {Noah}\ \bibnamefont
  {Finkelstein}}, \ and\ \bibinfo {author} {\bibfnamefont {H.~J.}\ \bibnamefont
  {Lewandowski}},\ }\bibfield  {title} {\enquote {\bibinfo {title} {Development
  and validation of the colorado learning attitudes about science survey for
  experimental physics},}\ }in\ \href@noop {} {\emph {\bibinfo {booktitle}
  {Physics Education Research Conference 2012}}},\ \bibinfo {series} {PER
  Conference}, Vol.\ \bibinfo {volume} {1513}\ (\bibinfo {address}
  {Philadelphia, PA},\ \bibinfo {year} {2012})\ pp.\ \bibinfo {pages}
  {442--445}\BibitemShut {NoStop}%
\bibitem [{\citenamefont {Zwickl}\ \emph {et~al.}(2014)\citenamefont {Zwickl},
  \citenamefont {Hirokawa}, \citenamefont {Finkelstein},\ and\ \citenamefont
  {Lewandowski}}]{zwickl2014eclass}%
  \BibitemOpen
  \bibfield  {author} {\bibinfo {author} {\bibfnamefont {Benjamin~M}\
  \bibnamefont {Zwickl}}, \bibinfo {author} {\bibfnamefont {Takako}\
  \bibnamefont {Hirokawa}}, \bibinfo {author} {\bibfnamefont {Noah}\
  \bibnamefont {Finkelstein}}, \ and\ \bibinfo {author} {\bibfnamefont
  {HJ}~\bibnamefont {Lewandowski}},\ }\bibfield  {title} {\enquote {\bibinfo
  {title} {Epistemology and expectations survey about experimental physics:
  Development and initial results},}\ }\href@noop {} {\bibfield  {journal}
  {\bibinfo  {journal} {Physical Review Special Topics-Physics Education
  Research}\ }\textbf {\bibinfo {volume} {10}},\ \bibinfo {pages} {010120}
  (\bibinfo {year} {2014})}\BibitemShut {NoStop}%
\bibitem [{\citenamefont {Wilcox}\ and\ \citenamefont
  {Lewandowski}(2016{\natexlab{a}})}]{wilcox2016eclass}%
  \BibitemOpen
  \bibfield  {author} {\bibinfo {author} {\bibfnamefont {Bethany~R.}\
  \bibnamefont {Wilcox}}\ and\ \bibinfo {author} {\bibfnamefont {H.~J.}\
  \bibnamefont {Lewandowski}},\ }\bibfield  {title} {\enquote {\bibinfo {title}
  {Students' epistemologies about experimental physics: Validating the colorado
  learning attitudes about science survey for experimental physics},}\ }\href
  {\doibase 10.1103/PhysRevPhysEducRes.12.010123} {\bibfield  {journal}
  {\bibinfo  {journal} {Phys. Rev. Phys. Educ. Res.}\ }\textbf {\bibinfo
  {volume} {12}},\ \bibinfo {pages} {010123} (\bibinfo {year}
  {2016}{\natexlab{a}})}\BibitemShut {NoStop}%
\bibitem [{\citenamefont {Wilcox}\ \emph {et~al.}(2016)\citenamefont {Wilcox},
  \citenamefont {Zwickl}, \citenamefont {Hobbs}, \citenamefont {Aiken},
  \citenamefont {Welch},\ and\ \citenamefont {Lewandowski}}]{wilcox2016admin}%
  \BibitemOpen
  \bibfield  {author} {\bibinfo {author} {\bibfnamefont {Bethany~R.}\
  \bibnamefont {Wilcox}}, \bibinfo {author} {\bibfnamefont {Benjamin~M.}\
  \bibnamefont {Zwickl}}, \bibinfo {author} {\bibfnamefont {Robert~D.}\
  \bibnamefont {Hobbs}}, \bibinfo {author} {\bibfnamefont {John~M.}\
  \bibnamefont {Aiken}}, \bibinfo {author} {\bibfnamefont {Nathan~M.}\
  \bibnamefont {Welch}}, \ and\ \bibinfo {author} {\bibfnamefont {H.~J.}\
  \bibnamefont {Lewandowski}},\ }\bibfield  {title} {\enquote {\bibinfo {title}
  {Alternative model for administration and analysis of research-based
  assessments},}\ }\href {\doibase 10.1103/PhysRevPhysEducRes.12.010139}
  {\bibfield  {journal} {\bibinfo  {journal} {Phys. Rev. Phys. Educ. Res.}\
  }\textbf {\bibinfo {volume} {12}},\ \bibinfo {pages} {010139} (\bibinfo
  {year} {2016})}\BibitemShut {NoStop}%
\bibitem [{\citenamefont {Wilcox}\ and\ \citenamefont
  {Lewandowski}(2016{\natexlab{b}})}]{wilcox2016gender}%
  \BibitemOpen
  \bibfield  {author} {\bibinfo {author} {\bibfnamefont {Bethany~R.}\
  \bibnamefont {Wilcox}}\ and\ \bibinfo {author} {\bibfnamefont {H.J.}\
  \bibnamefont {Lewandowski}},\ }\bibfield  {title} {\enquote {\bibinfo {title}
  {Surveying students' epistemologies about experimental physics: When is
  gender a factor?}}\ }\href@noop {} {\bibfield  {journal} {\bibinfo  {journal}
  {Phys. Rev. ST Phys. Educ. Res.}\ }\textbf {\bibinfo {volume} {under review}}
  (\bibinfo {year} {2016}{\natexlab{b}})}\BibitemShut {NoStop}%
\bibitem [{\citenamefont {Wilcox}\ and\ \citenamefont
  {Lewandowski}()}]{wilcox2016pedagogy}%
  \BibitemOpen
  \bibfield  {author} {\bibinfo {author} {\bibfnamefont {Bethany}\ \bibnamefont
  {Wilcox}}\ and\ \bibinfo {author} {\bibfnamefont {H.~J.}\ \bibnamefont
  {Lewandowski}},\ }\bibfield  {title} {\enquote {\bibinfo {title} {Impact of
  instructional approach on students' epistemologies about experimental
  physics},}\ }\bibfield  {booktitle} {\emph {\bibinfo {booktitle} {Physics
  Education Research Conference 2016}},\ }\href@noop {} {\ ,\ \bibinfo {pages}
  {Submitted}}\BibitemShut {NoStop}%
\bibitem [{\citenamefont {Mann}\ and\ \citenamefont
  {Whitney}(1947)}]{mann1947mwu}%
  \BibitemOpen
  \bibfield  {author} {\bibinfo {author} {\bibfnamefont {Henry~B}\ \bibnamefont
  {Mann}}\ and\ \bibinfo {author} {\bibfnamefont {Donald~R}\ \bibnamefont
  {Whitney}},\ }\bibfield  {title} {\enquote {\bibinfo {title} {On a test of
  whether one of two random variables is stochastically larger than the
  other},}\ }\href@noop {} {\bibfield  {journal} {\bibinfo  {journal} {The
  annals of mathematical statistics}\ ,\ \bibinfo {pages} {50--60}} (\bibinfo
  {year} {1947})}\BibitemShut {NoStop}%
\bibitem [{\citenamefont {Cohen}(1988)}]{cohen1988d}%
  \BibitemOpen
  \bibfield  {author} {\bibinfo {author} {\bibfnamefont {J.}~\bibnamefont
  {Cohen}},\ }\href {https://books.google.com/books?id=Tl0N2lRAO9oC} {\emph
  {\bibinfo {title} {Statistical Power Analysis for the Behavioral Sciences}}}\
  (\bibinfo  {publisher} {L. Erlbaum Associates},\ \bibinfo {year}
  {1988})\BibitemShut {NoStop}%
\bibitem [{\citenamefont {Rodriguez}\ \emph {et~al.}(2012)\citenamefont
  {Rodriguez}, \citenamefont {Brewe}, \citenamefont {Sawtelle},\ and\
  \citenamefont {Kramer}}]{rodriguez2012equity}%
  \BibitemOpen
  \bibfield  {author} {\bibinfo {author} {\bibfnamefont {Idaykis}\ \bibnamefont
  {Rodriguez}}, \bibinfo {author} {\bibfnamefont {Eric}\ \bibnamefont {Brewe}},
  \bibinfo {author} {\bibfnamefont {Vashti}\ \bibnamefont {Sawtelle}}, \ and\
  \bibinfo {author} {\bibfnamefont {Laird~H.}\ \bibnamefont {Kramer}},\
  }\bibfield  {title} {\enquote {\bibinfo {title} {Impact of equity models and
  statistical measures on interpretations of educational reform},}\ }\href
  {\doibase 10.1103/PhysRevSTPER.8.020103} {\bibfield  {journal} {\bibinfo
  {journal} {Phys. Rev. ST Phys. Educ. Res.}\ }\textbf {\bibinfo {volume}
  {8}},\ \bibinfo {pages} {020103} (\bibinfo {year} {2012})}\BibitemShut
  {NoStop}%
\bibitem [{\citenamefont {Holm}(1979)}]{holm1979hb}%
  \BibitemOpen
  \bibfield  {author} {\bibinfo {author} {\bibfnamefont {Sture}\ \bibnamefont
  {Holm}},\ }\bibfield  {title} {\enquote {\bibinfo {title} {A simple
  sequentially rejective multiple test procedure},}\ }\href@noop {} {\bibfield
  {journal} {\bibinfo  {journal} {Scandinavian journal of statistics}\ ,\
  \bibinfo {pages} {65--70}} (\bibinfo {year} {1979})}\BibitemShut {NoStop}%
\bibitem [{\citenamefont {{tinyurl.com/ECLASS-physics}}(2015)}]{ECLASSwebsite}%
  \BibitemOpen
  \bibfield  {author} {\bibinfo {author} {\bibnamefont
  {{tinyurl.com/ECLASS-physics}}},\ }\href@noop {} {} (\bibinfo {year}
  {2015})\BibitemShut {NoStop}%
\bibitem [{\citenamefont {Wildt}\ and\ \citenamefont
  {Ahtola}(1978)}]{wildt1978ancova}%
  \BibitemOpen
  \bibfield  {author} {\bibinfo {author} {\bibfnamefont {Albert~R}\
  \bibnamefont {Wildt}}\ and\ \bibinfo {author} {\bibfnamefont {Olli}\
  \bibnamefont {Ahtola}},\ }\href@noop {} {\emph {\bibinfo {title} {Analysis of
  covariance}}},\ Vol.~\bibinfo {volume} {12}\ (\bibinfo  {publisher} {Sage},\
  \bibinfo {year} {1978})\BibitemShut {NoStop}%
\bibitem [{\citenamefont {Day}\ \emph {et~al.}(2016)\citenamefont {Day},
  \citenamefont {Stang}, \citenamefont {Holmes}, \citenamefont {Kumar},\ and\
  \citenamefont {Bonn}}]{day2016gender}%
  \BibitemOpen
  \bibfield  {author} {\bibinfo {author} {\bibfnamefont {James}\ \bibnamefont
  {Day}}, \bibinfo {author} {\bibfnamefont {Jared~B.}\ \bibnamefont {Stang}},
  \bibinfo {author} {\bibfnamefont {N.~G.}\ \bibnamefont {Holmes}}, \bibinfo
  {author} {\bibfnamefont {Dhaneesh}\ \bibnamefont {Kumar}}, \ and\ \bibinfo
  {author} {\bibfnamefont {D.~A.}\ \bibnamefont {Bonn}},\ }\bibfield  {title}
  {\enquote {\bibinfo {title} {Gender gaps and gendered action in a first-year
  physics laboratory},}\ }\href@noop {} {\bibfield  {journal} {\bibinfo
  {journal} {Phys. Rev. ST Phys. Educ. Res.}\ }\textbf {\bibinfo {volume}
  {Accepted}} (\bibinfo {year} {2016})}\BibitemShut {NoStop}%
\bibitem [{\citenamefont {Miller}\ and\ \citenamefont
  {Chapman}(2001)}]{miller2001ancova}%
  \BibitemOpen
  \bibfield  {author} {\bibinfo {author} {\bibfnamefont {Gregory~A}\
  \bibnamefont {Miller}}\ and\ \bibinfo {author} {\bibfnamefont {Jean~P}\
  \bibnamefont {Chapman}},\ }\bibfield  {title} {\enquote {\bibinfo {title}
  {Misunderstanding analysis of covariance.}}\ }\href@noop {} {\bibfield
  {journal} {\bibinfo  {journal} {Journal of abnormal psychology}\ }\textbf
  {\bibinfo {volume} {110}},\ \bibinfo {pages} {40} (\bibinfo {year}
  {2001})}\BibitemShut {NoStop}%
\end{thebibliography}%

\end{document}